\documentclass[12pt,a4paper,aps,preprint,superscriptaddress,nofootinbib]{revtex4-1}
\usepackage[utf8]{inputenc}
\usepackage{graphicx}
\usepackage{amssymb}
\usepackage{textcomp}
\usepackage{amsmath}
\usepackage{tabularx}
\usepackage{bm}
\usepackage{times}
\usepackage{color}
\usepackage{slashed}
\usepackage{multirow}
\usepackage{verbatim}
\usepackage{cancel}
\usepackage{subfigure}
\usepackage{ulem}

\usepackage[colorlinks=true, pdfstartview=FitV, linkcolor=blue, citecolor=blue, urlcolor=blue]{hyperref}
\allowdisplaybreaks[4]

\def\lsim{\mathrel{\raise.3ex\hbox{$<$\kern-.75em\lower1ex\hbox{$\sim$}}}}
\def\gsim{\mathrel{\raise.3ex\hbox{$>$\kern-.75em\lower1ex\hbox{$\sim$}}}}

\newcommand{\GeV}{{\rm GeV}}
\newcommand{\TeV}{{\rm TeV}}

\definecolor{orange}{rgb}{1,0.5,0}

\begin{document}

\title{The electroweak magnetic monopole in the presence of KSVZ axion}

\author{Tong Li}
\email{litong@nankai.edu.cn}
\affiliation{
School of Physics, Nankai University, Tianjin 300071, China
}
\author{Rui-Jia Zhang}
\email{zhangruijia@mail.nankai.edu.cn}
\affiliation{
School of Physics, Nankai University, Tianjin 300071, China
}

\begin{abstract}
The Witten effect implies the dynamics of axion and magnetic monopole. The Cho-Maison monopole is a realistic electroweak monopole arisen in the Weinberg-Salam theory. This monopole of TeV scale mass motivates the dedicated search for electroweak monopole at colliders.
In this work we investigate the implication of KSVZ axion for the electroweak magnetic monopole. We use the spherically symmetric ansatz for the electroweak dyon and introduce the spherically symmetric function for the axion field. The effective Lagrangian is then shown in terms of the electroweak monopole part, the axion kinetic energy as well as the axion interaction term. We derive the consequent equations of motion in the presence of the axion-photon coupling and show the numerical results of the topological solutions. We then calculate the changed characteristics of the electroweak monopole such as the monopole mass and the electromagnetic charges, as well as the axion potential energy.
\end{abstract}


\maketitle


\section{Introduction}
\label{sec:Intro}

The axion and magnetic monopole are two of the longstanding and interesting candidates of physics beyond the standard model (SM). The pseudo-Goldstone boson axion was proposed to solve the strong CP problem in quantum chromodynamics (QCD), as a result of the spontaneous breaking of a QCD anomalous Peccei-Quinn (PQ) global symmetry $U(1)_{\rm PQ}$~\cite{Peccei:1977hh,Peccei:1977ur,Weinberg:1977ma,Wilczek:1977pj}. The chiral transformation of the quark fields with PQ charges leads to the anomaly under quantum
electrodynamics (QED) and the coupling between the axion field $a$ and electromagnetic fields
\begin{eqnarray}
{1\over 4}g_{a\gamma\gamma}a F_{\mu\nu}\tilde{F}^{\mu\nu}=-g_{a\gamma\gamma}a\vec{E}\cdot \vec{B}\;.
\end{eqnarray}
The QCD axion and axion-like particles (ALPs) can also play as dark matter through the misalignment mechanism~\cite{Preskill:1982cy,Dine:1982ah}. The theory of axion and the detection of axion couplings have received a wide interest in both theoretical and experimental
aspects (see Refs.~\cite{DiLuzio:2020wdo,Sikivie_2021} for recent reviews).

P.~Dirac first suggested the existence of
magnetic monopole in quantum theory in 1931~\cite{Dirac:1931kp}. It can be constructed from a non-singular Abelian theory~\cite{Wu:1975es} or can arise from the
spontaneous breaking of non-Abelian gauge symmetries, resulting to monopole in grand unification theory (GUT)~\cite{tHooft:1974kcl,Polyakov:1974ek}.
Besides GUT monopole, the existence of electroweak monopole in particular gains much interest and attention~\cite{Cho:1996qd,Hung:2020vuo,Alexandre:2019iub,Ellis:2016glu,Lazarides:2021los}. Cho and Maison showed that the Weinberg-Salam model has a topology of magnetic monopole and there does exist a new type of dyon solutions in the SM~\cite{Cho:1996qd} (see Ref.~\cite{Cho:2019vzo} for a recent review). This electroweak generalization of the Dirac monopole comes from the non-trivial topology of the SM. The $SU(2)$ of the Weinberg-Salam model has the same non-Abelian monopole topology as the Georgi-Glashow model. The $U(1)_Y$ part has the Abelian monopole topology. The Cho-Maison monopole (CMM) is thus a hybrid of 't Hooft-Polyakov monopole and Dirac monopole. Its magnetic charge is twice that of the Dirac monopole and the monopole mass is of the order of
\begin{eqnarray}
{4\pi\over e^2} M_W\simeq 11~{\rm TeV}\;,
\end{eqnarray}
where $e$ is the elementary electric charge and $M_W$ is the $W$ boson mass.
This low mass motives the detection of such realistic monopole at the ATLAS~\cite{ATLAS:2019wkg} or MoEDAL~\cite{MoEDAL:2019ort} detector of LHC or future high-energy upgrades. The electroweak monopole also leads to various impacts on cosmology, such as the primordial magnetic black holes and dark matter.

In 1979, E.~Witten pointed out that a CP violating term in the non-Abelian $SO(3)$ theory provides an additional electric charge for the ’t Hooft-Polyakov monopoles in this theory~\cite{Witten:1979ey}. The generic value of electric charge $q_e$ becomes
\begin{eqnarray}
q_e = e \Big(n_e+{\theta\over 2\pi}n_m\Big)\;,
\end{eqnarray}
where $n_e~(n_m)\in \mathbb{Z}$ is the electric (magnetic) quantum number, and $\theta$ denotes a CP violating parameter.
When substituting the parameter $\theta$ by the axion field, this so-called Witten effect built the close relationship between axion and magnetic monopole. W.~Fischler et al. then derived the dyon-axion dynamics under the classical electromagnetism in 1983~\cite{Fischler:1983sc}. There are many recent studies discussing the possible modification of standard axion electrodynamics inspired by Witten effect in both theory~\cite{Sokolov:2022fvs,Sokolov:2023pos,Heidenreich:2023pbi} and phenomenology~\cite{Li:2022oel,Tobar:2022rko,McAllister:2022ibe,Li:2023kfh,Li:2023aow,Tobar:2023rga,Patkos:2023lof,Dai:2024dkr}. There also exist some open questions in the axion-magnetic monopole aspect~\cite{Agrawal:2022yvu}.

What is then the implication of axion for the electroweak monopole? In this work we try to make the first attempt. We explore the solution of Cho-Maison electroweak monopole in the benchmark model of axion, i.e. the Kim-Shifman-Vainshtein-Zakharov (KSVZ) model~\cite{Kim:1979if,Shifman:1979if}. Besides axion interactions at low energies, the content of KSVZ model is exactly the same as the SM. We are able to construct the topological solution of the electroweak monopole in the SM. More importantly, inspired by the Witten effect, the characteristic features of electroweak monopole would be changed in light of the interaction of axion and electromagnetic field. For instance, the axion-photon coupling would induce additional electric charge for the monopole and modify the monopole mass.
The monopole background would also confine the axion potential energy. We will concentrate on the solution of spherically symmetrical function of axion field in a background magnetic field. The contributions of axion to the properties of electroweak monopole will be studied in details. These changes have impacts on both the existence of axion and the search for electroweak monopole.

This paper is organized as follows. In Sec.~\ref{sec:CMMandKSVZ} we overview the benchmark axion model and the electroweak monopole called Cho-Maison monopole. The topological solutions of electroweak monopole and axion field will be explored in Sec.~\ref{sec:solution}. In Sec.~\ref{sec:property} we discuss the impact of axion on the properties of electroweak monopole. Our conclusions are drawn in Sec.~\ref{sec:Con}.

\section{The KSVZ axion model and the electroweak monopole}
\label{sec:CMMandKSVZ}

We first overview the KSVZ axion model and the electroweak monopole in the SM called Cho-Maison monopole.
The Lagrangian for the KSVZ model is
\begin{eqnarray}
\mathcal{L}_{\rm KSVZ}=- \frac{1}{4}W_{\mu\nu}^a W^{a\mu\nu} - \frac{1}{4} B_{\mu\nu} B^{\mu\nu} + | D_\mu \Phi |^2 + |\partial_\mu\phi|^2 +\mathcal{L}_{\rm VLF}-V_{\rm KSVZ}(\Phi,\phi)\;,
\end{eqnarray}
where $W_{\mu\nu}^a\ (a=1,2,3)$ and $B_{\mu\nu}$ respectively denote the field strength tensors of $SU(2)_L$ and $U(1)_Y$ gauge fields, $\Phi$ is the SM Higgs doublet, and $\phi$ is a complex scalar singlet.
The covariant derivative to the Higgs doublet is
\begin{eqnarray}
D_\mu \Phi &=& ( \partial_\mu + \frac{i g}{2} \sigma^a W_\mu^a + \frac{i g^\prime}{2}  B_\mu  ) \Phi \,,
\end{eqnarray}
where $\sigma^a$ denotes the Pauli matrix, and $g$ and $g'$ are the gauge couplings of $SU(2)_L$ and $U(1)_Y$, respectively.
The KSVZ model introduces the kinetic term and Yukawa term for a vector-like fermion (VLF) in the fundamental of $SU(3)_c$, singlet under $SU(2)_L$, and neutral under hypercharge: $\mathcal{Q}\sim (3,1,0)$
\begin{eqnarray}
\mathcal{L}_{\rm VLF}=\overline{\mathcal{Q}}i\cancel{D}\mathcal{Q}-y_{\mathcal{Q}}\overline{\mathcal{Q}_L} \mathcal{Q}_R \phi +h.c.\;,
\end{eqnarray}
which emerges a $U(1)_{\rm PQ}$ symmetry and the scalar potential
\begin{eqnarray}
V_{\rm KSVZ}(\Phi,\phi)=\lambda_\Phi\Big(|\Phi|^2-\frac{v^2}{2}\Big)^2+\lambda_\phi\Big(|\phi|^2-\frac{v_a^2}{2}\Big)^2\;.
\end{eqnarray}
Here $v$ is the vacuum expectation value (vev) of the Higgs field $\Phi$ and the $U(1)_{\rm PQ}$ symmetry is spontaneously broken with the vev parameter $v_a$. The complex scalar field is decomposed as
\begin{eqnarray}
\phi={1\over \sqrt{2}} (v_a+\sigma_a)e^{ia/v_a}\;,
\end{eqnarray}
where $a$ denotes the axion field and $\sigma_a$ is the radial mode.
After performing the axial rotation of the VLF field
\begin{eqnarray}
\mathcal{Q} \to e^{-i\gamma_5 {a\over 2v_a}}\mathcal{Q}\;,
\end{eqnarray}
and removing the axion field in the Yukawa term, one obtains the following anomalous Lagrangian
\begin{eqnarray}
\delta\mathcal{L}_{\rm KSVZ}={\alpha_s N\over 4\pi} {a\over v_a} G^a_{\mu\nu}\tilde{G}^{a\mu\nu}+{\alpha E\over 4\pi} {a\over v_a} F_{\mu\nu}\tilde{F}^{\mu\nu}\;,
\end{eqnarray}
where $G_{\mu\nu}^a\ (a=1,\cdots,8)$ and $F_{\mu\nu}$ are the field strength tensors of $SU(3)_c$ and $U(1)_{\rm em}$, respectively, the dual field strengths are defined as $\tilde{X}_{\mu\nu}\equiv {1\over 2}\epsilon_{\mu\nu\alpha\beta}X^{\alpha\beta}$ with $\epsilon_{0123}=1$~\footnote{We follow the convention in Ref.~\cite{DiLuzio:2020wdo} and adopt the metric tensor as diag$(+1,-1,-1,-1)$ throughout the paper.}, and $E$ and $N=N_{\rm DW}/2$ are the anomaly coefficients with $N_{\rm DW}$ being the domain wall (DW) number~\footnote{DW number refers to the number of distinct minima in the potential of the axion field. If the axion potential has $N_{\rm DW}$ distinct minima, $N_{\rm DW}$ domain walls can form when the axion field settles into different vacua in different space regions. The $N_{\rm DW}=1$ model is taken throughout the paper.}. Throughout the paper, we only consider the second coupling between axion and the physical photon. By taking into account both the quark kinetic term and the quark mass operator, the axion-photon coupling can be given as
\begin{eqnarray}
\mathcal{L}_\text{axion~int.}=(1/4)\,g_{a\gamma\gamma} a F_{\mu\nu}\tilde{F}^{\mu\nu}
\label{eq:aphoton}
\end{eqnarray}
with
\begin{eqnarray}
g_{a\gamma\gamma}\equiv\frac{c_{a\gamma\gamma}}{v_a}={\alpha\over \pi v_a} \Big[E-N\Big({2\over 3}{4m_d+m_u\over m_u+m_d}\Big)\Big]\simeq {\alpha\over \pi v_a}(E-1.92N)\;.
\end{eqnarray}
We take $g_{a\gamma\gamma}\simeq -0.0024/v_a$ in the minimal KSVZ model. The astrophysical constraint in Ref.~\cite{Dolan:2022kul} gives $|g_{a\gamma\gamma}|\leq 0.34\times 10^{-10}~{\rm GeV}^{-1}$ which results in $v_a\geq 7.1\times 10^7~{\rm GeV}$ here. Below we take $v_a=10^{12}~{\rm GeV}$ as an illustrative value.

Next we show the ansatz for the spherically symmetry solution of the Cho-Maison monopole in the KSVZ model with the effective Lagrangian
\begin{eqnarray}
\mathcal{L}_{eff}&=&- \frac{1}{4}W_{\mu\nu}^a W^{a\mu\nu} - \frac{1}{4} B_{\mu\nu} B^{\mu\nu} + | D_\mu \Phi |^2 + |\partial_\mu\phi|^2 \nonumber \\
&&-\lambda_\Phi\Big(|\Phi|^2-\frac{v^2}{2}\Big)^2-\lambda_\phi\Big(|\phi|^2-\frac{v_a^2}{2}\Big)^2+{1\over 4}g_{a\gamma\gamma} a F_{\mu\nu}\tilde{F}^{\mu\nu}\;.
\label{eq:Leff}
\end{eqnarray}
The authors of Refs.~\cite{Cho:1996qd,Cho:2019vzo} make use of the gauge-independent Abelian decomposition to Abelianize the non-Abelian gauge theory. The Abelianized dual potential has both a non-topological electric potential and a topological magnetic potential. They verify that the Abelian decomposition of the Weinberg-Salam model is similar to that of the Georgi-Glashow model and they have the same $SU(2)$ monopole topology.
Inspired by this implication, one can parameterize the Higgs doublet in terms of the spherical coordinates $(t,r,\theta,\varphi)$ as~\cite{Cho:1996qd,Cho:2019vzo}
\begin{eqnarray}
\Phi&=& \frac{1}{ \sqrt{2} } \rho (r) \xi (\theta\,,\varphi)\,, \qquad \xi^\dag \xi =1\,,
\label{equ:higgs}
\end{eqnarray}
with $\rho(r)$ being a real function, and $\xi(\theta\,, \varphi )=i(\sin{\theta\over 2}e^{-i\varphi},-\cos{\theta\over 2})^T$ being complex unit doublet.
The ansatz for the spherically symmetry solution of the electroweak dyon is given as~\cite{Cho:1996qd,Cho:2019vzo}
\begin{eqnarray}
g \vec W_\mu &=& A(r) (\partial_\mu t) \hat r + ( f(r)-1) \hat r \times \partial_\mu \hat r \,,\nonumber\\
g^\prime B_\mu &=& B(r) \partial_\mu t - (1 -\cos\theta ) \partial_\mu \varphi\;,
\label{equ:CMM_solution}
\end{eqnarray}
with $\hat r = - \xi^\dag \vec \sigma \xi= (\sin\theta \cos \varphi\,, \sin\theta \sin\varphi\,, \cos\theta ) $.
Either using the unitary gauge or using the gauge independent Abelian decomposition, the physical fields are expressed as~\cite{Cho:1996qd,Cho:2019vzo}
\begin{eqnarray}
A_\mu &=& e\Big({A(r)\over g^2}+{B(r)\over g^{\prime 2}} \Big)\partial_\mu t -{1\over e}(1-\cos\theta)\partial_\mu \varphi\;,
\label{eq:Amu}\\
W_\mu&=&{if(r)\over \sqrt{2}g}e^{+i\varphi}( \partial_\mu \theta + i\sin\theta \partial_\mu \varphi)\;,\\
W_\mu^{\ast}&=&{if(r)\over \sqrt{2}g}e^{-i\varphi}(-\partial_\mu \theta + i\sin\theta \partial_\mu \varphi)\;,\\
Z_\mu&=&{e\over gg'} (A(r)-B(r)) \partial_\mu t\;,
\end{eqnarray}
where $W_\mu$ ($W_\mu^\ast$) denotes the normal mass eigenstate of gauge boson $W_\mu^+$ ($W_\mu^-$).

The complex scalar singlet is parameterized as
\begin{eqnarray}
\phi={1\over \sqrt{2}} (v_a+\sigma_a)e^{ia(r)/v_a}\;,
\label{equ:higgs_singlet}
\end{eqnarray}
where $\sigma_a$ as the radial component is supposed to be stabilized and will not be considered below, and $a(r)$ is the axion (angular) component.
The axion-photon coupling in Eq.~(\ref{eq:aphoton}) leads to the modified equations of motion (EoM)
\begin{eqnarray}
\partial_\mu (F^{\mu\nu} -g_{a\gamma\gamma} a \tilde{F}^{\mu\nu}) = 0 \;,
\end{eqnarray}
and the relevant modified Gauss's law for the electric field
\begin{eqnarray}
\vec{\nabla}\cdot (\vec{E} + g_{a\gamma\gamma}a \vec{B})&= 0\;.
\label{equ:modified_Maxwell}
\end{eqnarray}
Suppose there is a magnetic monopole with a magnetic charge $q_m$, the above Gauss's law changes the usual electric charge quantization $q_e/e=n\in \mathbb{Z}$ to
\begin{eqnarray}
{q_e\over e}+{q_m\over e}g_{a\gamma\gamma}a(\infty)=n\in \mathbb{Z}\;.
\end{eqnarray}
This turns out to be the generic feature of Witten effect in $U(1)$ gauge group.

\section{The solution of electroweak monopole in the presence of axion}
\label{sec:solution}

Using the spherically symmetric ansatz in Eqs.~(\ref{equ:higgs}) and (\ref{equ:CMM_solution}), we can rewrite the above effective Lagrangian in Eq.~(\ref{eq:Leff}) in terms of five radial functions $\rho(r)$, $f(r)$, $A(r)$, $B(r)$ and $a(r)$. The complete expansion of the Lagrangian contains three parts: the Cho-Maison monopole part, the axion kinetic energy and the axion interaction
\begin{eqnarray}
\mathcal{L}_{eff}=\mathcal{L}_{\rm CMM}+\mathcal{L}_{\rm axion~kin.}+\mathcal{L}_{\rm axion~int.}
\label{equ:eff}
\end{eqnarray}
with
\begin{eqnarray}
\mathcal{L}_{\rm CMM}&=&\frac{1}{2}(\partial_\mu\rho)^2-\frac{\lambda_\Phi}{4}(\rho^2-v^2)^2-\frac{1}{4}{F'}_{\mu\nu}^2-\frac{1}{4}B_{\mu\nu}^2+\frac{g^2}{4}\rho^2W_\mu^*W^\mu+\frac{g^2+{g'}^2}{8}\rho^2Z_\mu^2
\nonumber\\
&-&\frac{1}{2}\left\vert(D_\mu W_\nu-D_\nu W_\mu)+ie\frac{g}{g'}(Z_\mu W_\nu-Z_\nu W_\mu)\right\vert^2\nonumber\\
&+&ieF_{\mu\nu}{W^*}^\mu W^\nu+ie\frac{g}{g'}Z_{\mu\nu}{W^*}^\mu W^\nu+\frac{g^2}{4}(W_\mu^* W_\nu-W_\nu^*W_\mu)^2\nonumber\\
&=&\frac{1}{2g^2}\left(\frac{dA}{dr}\right)^2+\frac{1}{8}(A-B)^2\rho^2+\frac{f^2A^2}{g^2r^2}-\frac{f^2\rho^2}{4r^2}+\frac{1}{2g'^2}\left(\frac{dB}{dr}\right)^2-\frac{1}{g^2r^2}\left(\frac{df}{dr}\right)^2\nonumber \\
&&-\frac{1}{2g'^2r^4}-\frac{1}{2}\left(\frac{d\rho}{dr}\right)^2-\frac{\lambda_\Phi}{4}(\rho^2-v^2)^2-\frac{(f^2-1)^2}{2g^2r^4}\;,\\
\mathcal{L}_{\rm axion~kin.}&=&\frac{1}{2}(\partial_\mu a)^2=-\frac{1}{2}\left(\frac{da}{dr}\right)^2\;,\\
\mathcal{L}_{\rm axion~int.}&=&\frac{1}{4}g_{a\gamma\gamma}aF_{\mu\nu}\tilde{F}^{\mu\nu}=-g_{a\gamma\gamma}\frac{a}{r^2}\Big[\left(\frac{dA}{dr}\right)\left(\frac{\sin^2\theta_W}{g^2}(1-f^2)+\frac{\cos^2\theta_W}{g^2}\right)\nonumber \\
&&+\left(\frac{dB}{dr}\right)\left(\frac{\sin^2\theta_W}{g'^2}(1-f^2)+\frac{\cos^2\theta_W}{g'^2}\right)\Big]\;,
\end{eqnarray}
where ${F'}_{\mu\nu}=\partial_\mu W_\nu^3-\partial_\nu W_\mu^3=\frac{e}{g}F_{\mu\nu}+\frac{e}{g'}Z_{\mu\nu}$ with $Z_{\mu\nu}=\partial_\mu Z_\nu - \partial_\nu Z_\mu$, and $D_\mu=\partial_\mu+ie A_\mu$.
Note that we ignore the QCD axion potential here, as we focus on the impact of the axion-photon interaction on the electroweak monopole properties. This approximation is justified in the regime where the characteristic length scale of the axion configuration is much smaller than the axion Compton wavelength, i.e. $r\ll m_a^{-1}$. Consequently, the QCD axion potential-induced term $\frac{\partial V}{\partial a}\sim m_a^2 a$ is much smaller than the remaining terms in Eq.~\eqref{equ:EoM}, which scale as $1/r^n$. It is therefore consistent to neglect the axion potential when solving for the local field configuration around the monopole.
Then, the equations of motion for the above radial functions can be obtained as
\begin{eqnarray}
&&\frac{d^2\rho}{dr^2}+\frac{2}{r}\frac{d\rho}{dr}-\frac{f^2}{2r^2}\rho+\frac{1}{4}(A-B)^2\rho=\lambda_\Phi(\rho^2-v^2)\rho\;,
\label{eq:rho}\\
&&\frac{d^2f}{dr^2}-\frac{(f^2-1)f}{r^2}-T_f=(\frac{g^2\rho^2}{4}-A^2)f\;,
\label{eq:f}\\
&&\frac{d^2A}{dr^2}+\frac{2}{r}\frac{dA}{dr}-\frac{2f^2A}{r^2}+T_A=\frac{g^2}{4}(A-B)\rho^2\;,
\label{eq:A}\\
&&\frac{d^2B}{dr^2}+\frac{2}{r}\frac{dB}{dr}+T_B=-\frac{g'^2}{4}(A-B)\rho^2\;,
\label{eq:B}\\
&&\frac{d^2a}{dr^2}+\frac{2}{r}\frac{da}{dr}+T_a=0\;,
\label{equ:EoM}
\end{eqnarray}
where
\begin{eqnarray}
T_f&=&-g_{a\gamma\gamma}af\left[\sin^2\theta_W\left(\frac{dA}{dr}\right)+\cos^2\theta_W\left(\frac{dB}{dr}\right)\right]\;,\\
T_A&=&T_B=-g_{a\gamma\gamma}\frac{1}{r^2}\left[\left(\frac{da}{dr}-\frac{2a}{r}\right)\left(1-\sin^2\theta_W f^2\right)-2\sin^2\theta_Waf\left(\frac{df}{dr}\right)\right]\;,\\
T_a&=&-g_{a\gamma\gamma}\frac{1}{r^2}\Big[\left(\frac{dA}{dr}\right)\left({1\over g^2}-\frac{\sin^2\theta_W}{g^2}f^2\right)+\left(\frac{dB}{dr}\right)\left({1\over g^{\prime 2}}-\frac{\sin^2\theta_W}{g'^2}f^2\right)\Big]\;.
\label{equ:T}
\end{eqnarray}
When $g_{a\gamma\gamma}$ is absent, the first four equations of motion would restore to those for pure Cho-Maison monopole~\cite{Cho:2019vzo}. The last equation becomes the equation of motion for axion field.

As stated in Refs.~\cite{Cho:1996qd,Cho:2019vzo}, the energy of pure Cho-Maison monopole is infinite according to the Lagrangian $\mathcal{L}_{\rm CMM}$. To make the electroweak monopole finite, one has to include the quantum correction and regularize the Cho-Maison monopole~\cite{Bae:2002bm,Cho:2012bq,Cho:2013vba}. The ultraviolet regularization introduces the hypercharge $U(1)$ permittivity $\epsilon(\rho)$ and changes the CMM Lagrangian to
\begin{eqnarray}
{\mathcal{L}'}_{\rm CMM}&=&\frac{1}{2}(\partial_\mu\rho)^2-\frac{\lambda_\Phi}{4}(\rho^2-v^2)^2-\frac{1}{4}{F'}_{\mu\nu}^2-\epsilon(\rho)\frac{1}{4}B_{\mu\nu}^2+\frac{g^2}{4}\rho^2W_\mu^*W^\mu+\frac{g^2+{g'}^2}{8}\rho^2Z_\mu^2
\nonumber\\
&-&\frac{1}{2}\left\vert(D_\mu W_\nu-D_\nu W_\mu)+ie\frac{g}{g'}(Z_\mu W_\nu-Z_\nu W_\mu)\right\vert^2\nonumber\\
&+&ieF_{\mu\nu}{W^*}^\mu W^\nu+ie\frac{g}{g'}Z_{\mu\nu}{W^*}^\mu W^\nu+\frac{g^2}{4}(W_\mu^* W_\nu-W_\nu^*W_\mu)^2\nonumber\\
&=&\frac{1}{2g^2}\left(\frac{dA}{dr}\right)^2+\frac{1}{8}(A-B)^2\rho^2+\frac{f^2A^2}{g^2r^2}-\frac{f^2\rho^2}{4r^2}+\frac{\epsilon(\rho)}{2g'^2}\left(\frac{dB}{dr}\right)^2-\frac{1}{g^2r^2}\left(\frac{df}{dr}\right)^2\nonumber \\
&&-\frac{\epsilon(\rho)}{2g'^2r^4}-\frac{1}{2}\left(\frac{d\rho}{dr}\right)^2-\frac{\lambda_\Phi}{4}(\rho^2-v^2)^2-\frac{(f^2-1)^2}{2g^2r^4}\;.
\label{eq:LUV}
\end{eqnarray}
Then, the equation of motion in Eq.~(\ref{eq:rho}) and Eq.~(\ref{eq:B}) are modified as
\begin{eqnarray}
&&\frac{d^2\rho}{dr^2}+\frac{2}{r}\frac{d\rho}{dr}-\frac{f^2}{2r^2}\rho+\frac{1}{4}(A-B)^2\rho=\lambda_\Phi(\rho^2-v^2)\rho+\frac{\epsilon'}{2{g'}^2}\left(\frac{1}{r^4}-\Big({dB\over dr}\Big)^2\right)\;,\\
&&\frac{d^2B}{dr^2}+2\left(\frac{1}{r}+\frac{\epsilon'}{2\epsilon}{d\rho\over dr}\right)\frac{dB}{dr}+\frac{T_B}{\epsilon}=-\frac{g'^2}{4\epsilon}(A-B)\rho^2\;,
\end{eqnarray}
where $\epsilon'=d\epsilon/d\rho$. To make the monopole energy finite, the regular solution can be written as
\begin{eqnarray}
\epsilon(\rho)\simeq\left(\frac{\rho}{\rho_0}\right)^n\;,~~\rho(r)\simeq c_0r^{\delta_0}\;,
\label{eq:alphabeta}
\end{eqnarray}
where $\rho_0=v$ and $\delta_0$ depends on the choice of $n$. In particular, Ref.~\cite{Zhang:2020xsg} imposes constraints on the range of $n$ values. To obtain finite energy  monopoles, $n$ needs to be greater than 2.
In principle, the solutions of monopole (dyon) can exist for continuous values of $n$, giving rise to a family of localized monopole (dyon) solutions with different energies. Which value of $n$ is physically realized in nature remains an open question that must ultimately be constrained by experiment.
Under appropriate boundary conditions, we can numerically integrate the above differential equations to obtain the electroweak monopole solution with KSVZ axion.
Following Ref.~\cite{Cho:2019vzo}, we choose the following boundary conditions for UV regularized electroweak monopole
\begin{eqnarray}
&&\rho(0)=0\;,~\rho(\infty)=v\;,~~f(0)=1\;,~f(\infty)=0\;,\nonumber \\
&&A(0)=0\;,~B(0)=0\;,~A(\infty)=B(\infty)=g v/4=M_W/2\;,
\end{eqnarray}
for the Higgs boson, $W$ boson and $Z$ boson represented by $\rho$, $f$ and $A-B$, respectively.
As the axion is defined as an angular variable, we set the boundary conditions for axion as~\cite{Fischler:1983sc,Marsh:2015xka}
\begin{eqnarray}
a(0)=0,~a(\infty)=\theta_\infty v_a\;,
\label{eq:axion_boundary}
\end{eqnarray}
where the angle $\theta_\infty$ refers to a dimensionless parameter characterizing the asymptotic axion background. Here, $a(\infty)$ is treated as an external boundary condition for the local monopole induced axion profile. A possible cosmological interpretation of this asymptotic background is that it may originate from the standard misalignment mechanism on large scales. In such case one may parametrically identify $\theta_\infty$ with the average misalignment angle $\theta_{\rm mis}$. The corresponding axion relic abundance is approximately given by~\cite{Fox:2004kb}
\begin{eqnarray}
\Omega_a h^2\sim 2\times 10^{4} \Big({v_a\over 10^{16}~{\rm GeV}}\Big)^{7/6} \theta_\text{mis}^2 \simeq 0.12\;.
\end{eqnarray}
Although the analysis carried out here treats $a(\infty)$ simply as a fixed external boundary value, in the numerical calculations below we continue to adopt the cosmologically motivated choice $a(\infty)=\theta_{\rm mis}v_a$ as a benchmark boundary condition.

\begin{figure}[htbp]
\begin{center}
\includegraphics[width=0.4\linewidth]{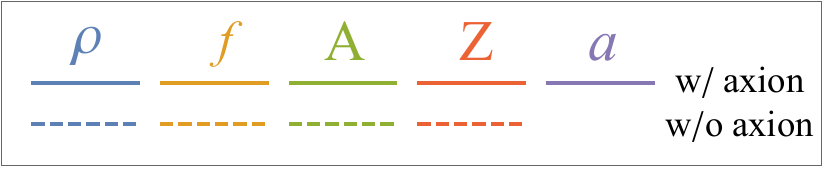}\\
\includegraphics[scale=1,width=0.49\linewidth]{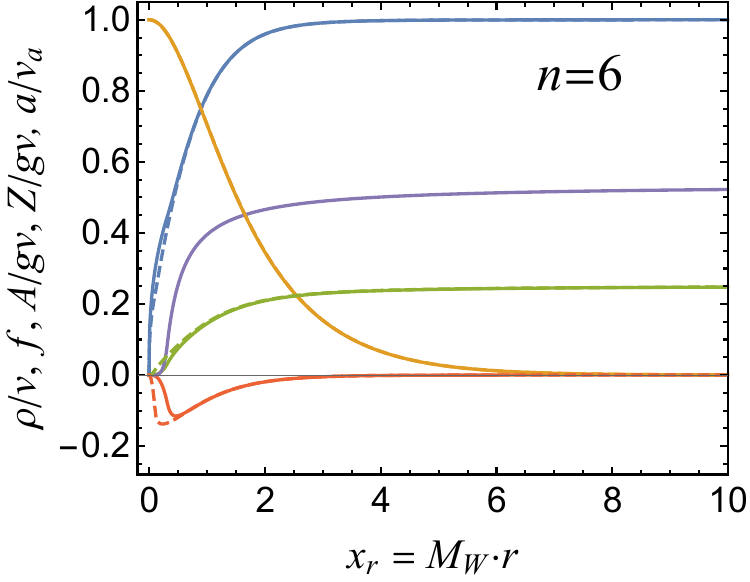}
\includegraphics[scale=1,width=0.49\linewidth]{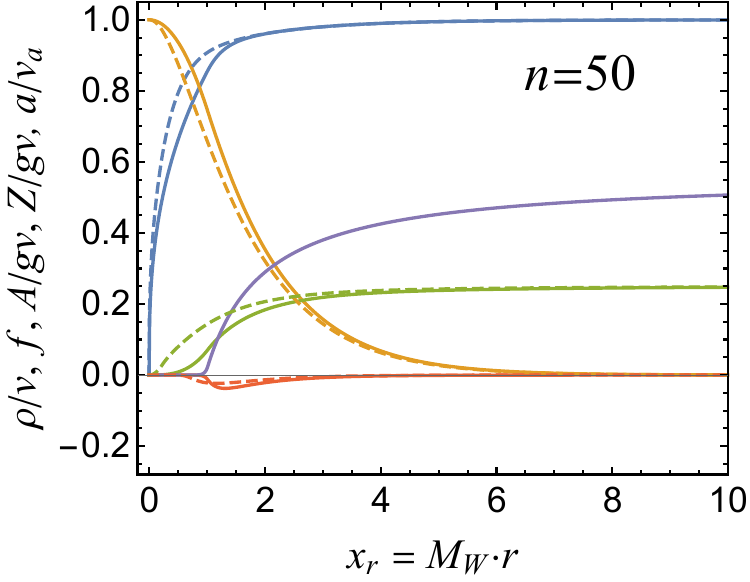}
\end{center}
\caption{The finite-energy dyon solutions in the presence of the KSVZ axion (solid lines) with those in its absence (dashed lines) for $n=6$ (left panel) and $n=50$ (right panel), respectively.
The anomaly coefficients of axion in KSVZ model are $E=0$ and $N=1/2$. The PQ scale is set as $v_a=10^{12}~\GeV$ for illustration.
}
\label{fig:axion-dyon}
\end{figure}

For above differential equations in Eqs.~(\ref{eq:rho},\ref{eq:f},\ref{eq:A},\ref{eq:B},\ref{equ:EoM}), we use the relaxation method to numerically get iterative solutions~\cite{Press2007} for the region of $x_r=M_W\cdot r>x_{\rm{min}}$. The specific value of $x_{\rm{min}}$ depends on the choice of $n$. A larger $n$ results in a larger $x_{\rm{min}}$, but it never exceeds a certain upper bound. This bound ensures the continuity of the first derivatives of all solutions and maintains the residuals of the differential equations below $10^{-3}$.
Besides, for the region of $x<x_{\rm{min}}$, we take the analytical approximation in each iteration as
\begin{eqnarray}
\rho\simeq c_0 r^{\delta_0},~~f\simeq 1+c_1 r^{\delta_1},~~A\simeq c_2 r^{\delta_2},~~Z\simeq b_0+c_3  r^{\delta_3},~~a\simeq c_4 r^{\delta_4}\;,
\end{eqnarray}
where the exponents $\delta_{0,1,2,3,4}$ must satisfy the following conditions with KSVZ axion
\begin{eqnarray}
    &&\delta_0=\begin{cases}
    \frac{2}{n-2}& \text{if } 2<n \leq 3\;,\\
\frac{\sqrt{3}-1}{2} & \text{if } n \geq 3.45\;,
\end{cases}\;\\
    &&\delta_1=2\;,\quad\delta_2\geq3\;,\\
    &&\delta_3\geq\begin{cases}
    n\left(\frac{2}{n-2}\right)-\sqrt{3}+1& \text{if } 2<n \leq 3\;,\\
 3 & \text{if } 3.45\leq n \leq 10.2\;, \\
(\frac{\sqrt{3}-1}{2})(n-2)  & \text{if } n > 10.2\;,
\end{cases}\\
&&\delta_4\geq\begin{cases}
    n\left(\frac{2}{n-2}\right)+3& \text{if } 2<n \leq 3\;,\\
\left(\frac{\sqrt{3}-1}{2}\right)n+3 & \text{if } n \geq 3.45\;.
\end{cases}\;
\end{eqnarray}
The $c_{0,1,2,3,4}$ coefficients are chosen to match the numerical solution smoothly.
In Fig.~\ref{fig:axion-dyon}, as illustration, we compare the finite energy dyon solutions in the presence of the KSVZ axion (solid lines) with those in its absence (dashed lines) for $n=6$ (left panel) and $n=50$ (right panel).
The anomaly coefficients of axion in KSVZ model are $E=0$ and $N=1/2$. The PQ scale is set as $v_a=10^{12}~{\rm GeV}$ for illustration. In the presence of small axion-photon coupling, the monopole solutions are compatible with those of pure Cho-Maison electroweak monopole in Ref.~\cite{Cho:2019vzo}. Nevertheless, the solution of axion field is correlated with functions $A(r)$ and $B(r)$ as seen in Eq.~(\ref{equ:EoM}), and its presence changes the dyon solutions for large $n$.

\begin{figure}[htbp]
\centering
\includegraphics[width=0.6\linewidth]{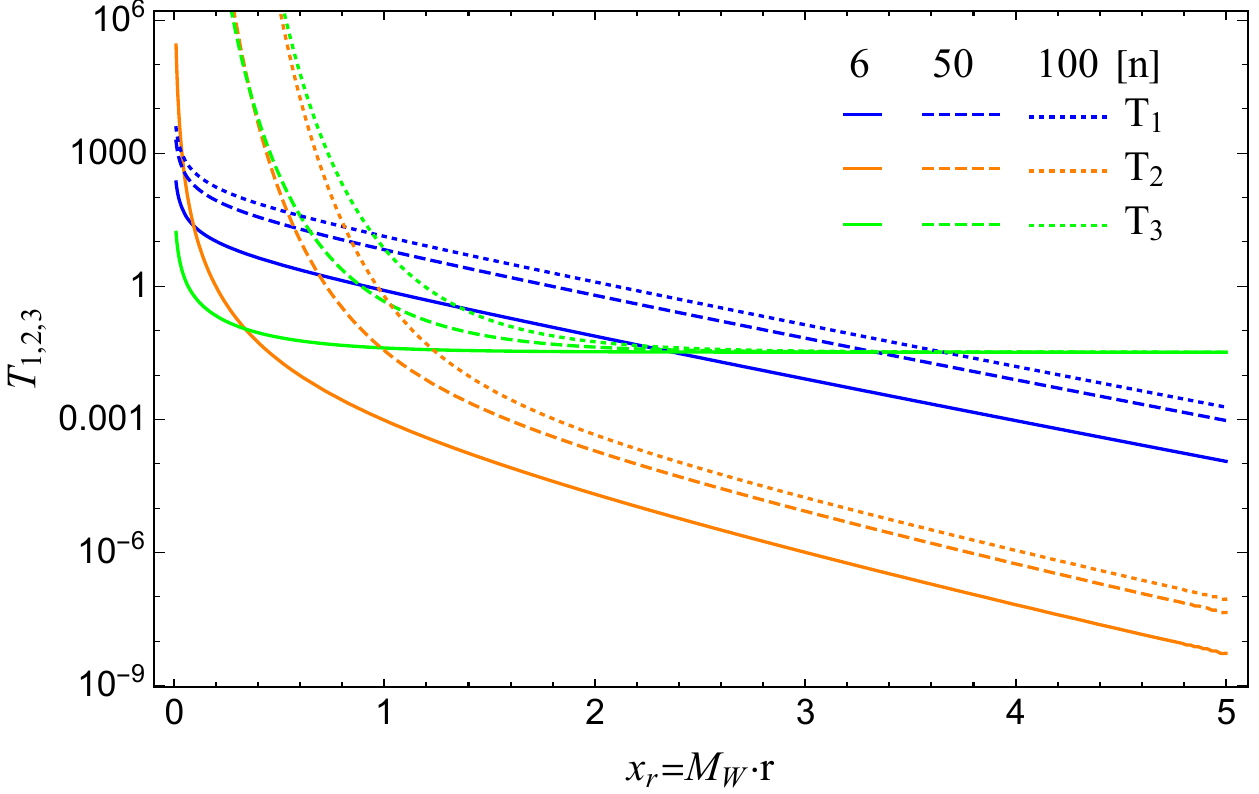}
\caption{Behavior of the dyon solutions' divergence as a function of $x_r=M_W\cdot r$ for different values of $n$. Three divergent terms are $T_1$ (blue), $T_2$ (orange), and $T_3$ (green) with represent values of $n$: $n = 6$ (solid), 50 (dashed), and 100 (dotted).
}
\label{fig:T123}
\end{figure}

In this regularization approach incorporating the permittivity-like function $\epsilon(\rho)=(\rho/\rho_0)^n$, the value of $n$ appears to have no upper bound--meaning it can, in principle, extend to infinity. From EoM, one can see that certain terms in the $\rho$ and $Z$ equations exhibit $n$-dependence. Specifically, in $Z$ equation:
 \begin{eqnarray}
    \frac{d^2Z}{dr^2}+\Big(\frac{2}{r}+T_{\epsilon\rho}\Big)\frac{dZ}{dr}-\frac{2f^2}{r^2}A -T_{\epsilon\rho A}+T_A\left(1-\frac{1}{\epsilon}\right)= \frac{1}{4}\left(g^2+\frac{g'^2}{\epsilon}\right)Z\rho^2\;,
\end{eqnarray}
the following three relevant terms with $y_0=\rho/\rho_0$
\begin{eqnarray}
    &&T_{\epsilon\rho A}=T_{\epsilon\rho}\times\frac{dA}{dr}=\frac{n}{\rho}\frac{d\rho}{dr}\frac{dA}{dr}\propto T_1=\frac{n}{y_0}\frac{dy_0}{dx}\;,\\
    &&T_A\left(1-\frac{1}{\epsilon}\right)\propto T_2=\frac{c_{a\gamma\gamma}}{x^2}\left(1-\frac{1}{\epsilon(x)}\right)\;,\\
    &&\frac{g'^2}{4\epsilon}Z\rho^2\propto T_3=\frac{g'^2}{4}\frac{1}{\epsilon(x)}y_0^2\;
\end{eqnarray}
are sensitive to the choice of $n$.
In particular, the latter two terms exhibit strong divergence when $n$ becomes sufficiently large, leading to a singularity in the $Z$ equation. To obtain a stable solution satisfying the core boundary condition $Z(0)=0$, the practical strategy is to suppress these two divergent contributions by pushing the solutions of $Z$ and $a$ to approximately zero in an inner core region. Fig.~\ref{fig:T123} shows their divergence patterns as functions of $x_r=M_W \cdot r$ for illustrative numbers of $n$. Since all $T$ terms are associated with the $y_0$ solution, to estimate the divergence of the $T_1$, $T_2$, and $T_3$ terms without first solving them for $y_0$, we approximate $y_0$ using the non-regularized solution of the SM dyon as a benchmark. The latter two terms related to
$(1-1/\epsilon)~(\text{term }T_2)$ and $1/\epsilon~(\text{term }T_3)$ dominate the singular behavior and grow much faster than $T_{\epsilon\rho A}~(\text{term}~T_1 )$.
As $n$ increases, the divergent region of these two terms extends outward from the origin. This means that the range over which $Z$ and $a$ are suppressed to zero also becomes larger.

\begin{figure}[htbp]
\begin{center}
\includegraphics[width=0.4\linewidth]{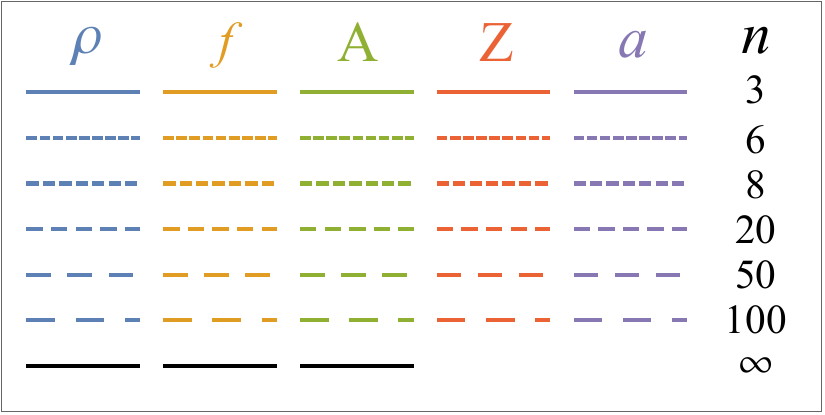}\\
\includegraphics[width=0.49\linewidth]{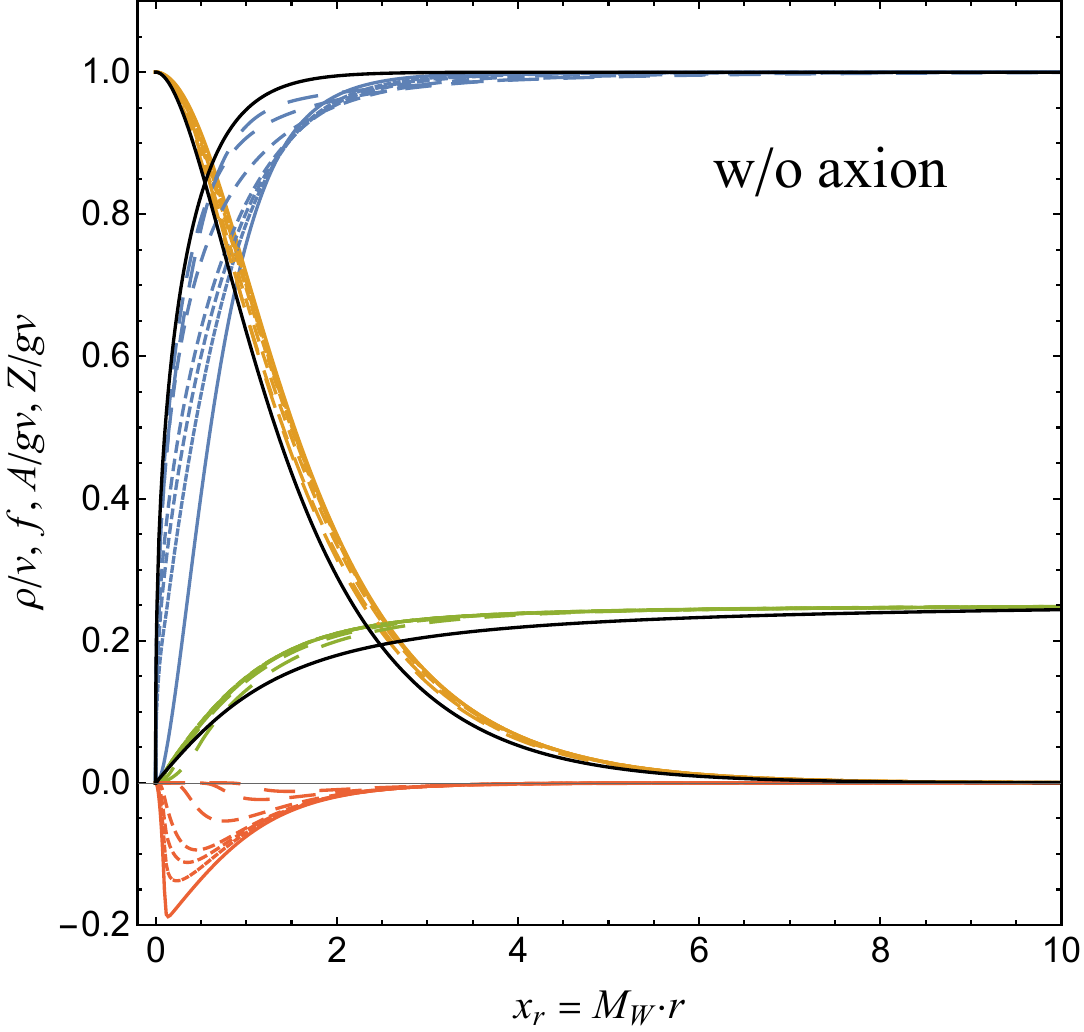}
\hfill
\includegraphics[width=0.49\linewidth]{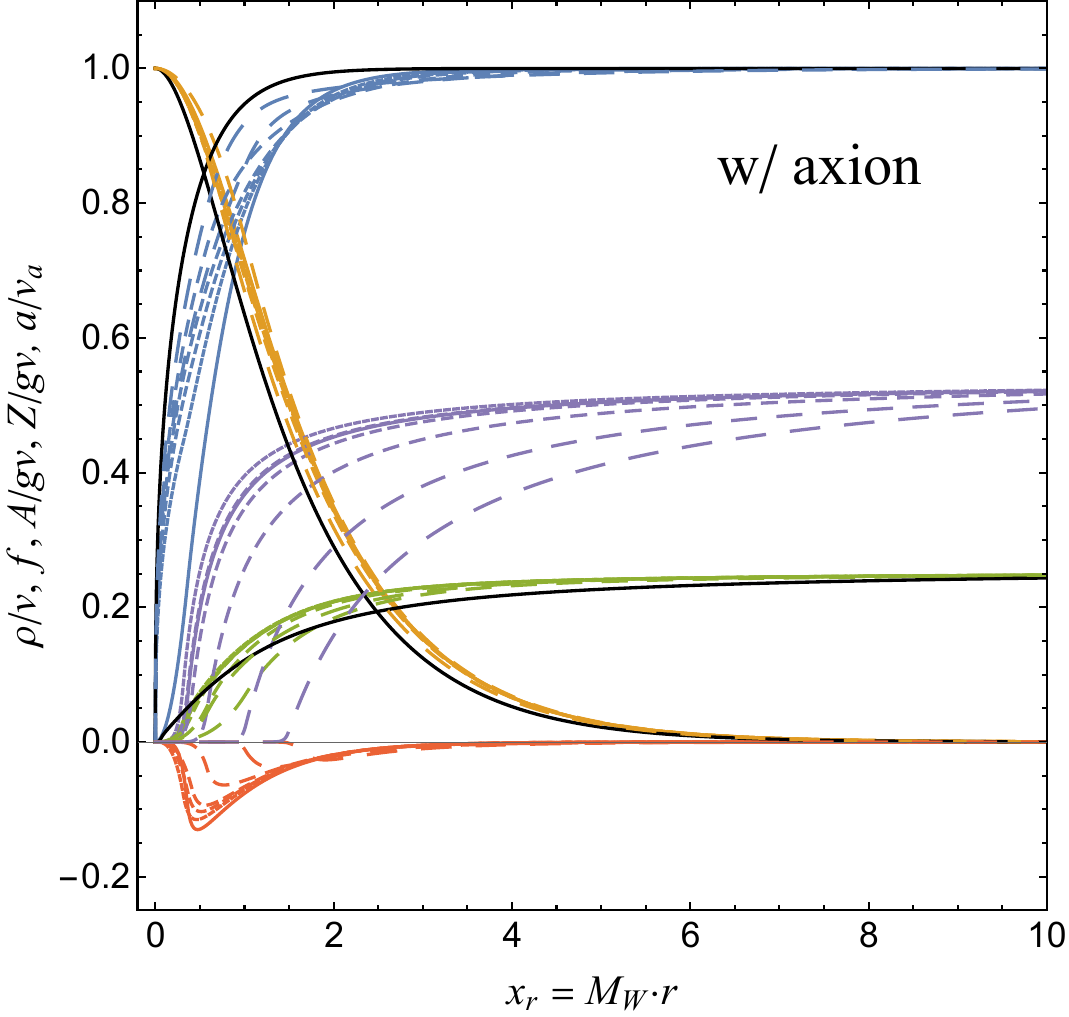}
\end{center}
\caption{The solutions for the SM dyon (left panel) and the dyon in presence of KSVZ axion (right panel) under different values of $n$.
The results corresponding to $n = 3$, 6, 8, 20, 50, and 100 are shown. The black solid line represents the solution for $n = \infty$.
}
\label{fig:dyon_sol}
\end{figure}

In Fig.~\ref{fig:dyon_sol}, we present the solutions of the SM dyon (left panel) and the dyon in presence of KSVZ axion (right panel) for different values of $n$.
The behavior shown in Fig.~\ref{fig:T123} manifests a wider suppression region for larger $n$.
In the limit $n \to \infty$, the divergent region covers the entire domain. To ensure a stable solution in this case, $Z$ and $a$ must be suppressed to zero everywhere, leaving only $\rho,~f,~A$ as non-zero solutions. In this case, $\rho,~f,~A$ decouple from the axion solution $a$, and then revert to the SM solutions (see the black curve in Fig.~\ref{fig:dyon_sol}).

We note that the suppression of the solution to zero is not unique to the dyon with KSVZ axion. In fact, for the electroweak dyon with vanishing axion field, i.e., $T_A=0$, a divergent term proportional to $T_3$ still remains. As shown in Fig.~\ref{fig:T123}, the divergence of this term (green) extends over an even wider range than $T_2$ term (orange). Consequently, the SM dyon also exhibits a suppression phenomenon as $n$ increases. However, in the absence of the $T_2$ divergent term, the overall divergence is less severe. Thus, the suppression ranges of $Z$ are correspondingly smaller than those with the KSVZ axion, as shown in the left panel of Fig.~\ref{fig:dyon_sol}.

\section{Characteristics of electroweak monopole and axion}
\label{sec:property}

\subsection{The monopole energy}

We first show the monopole energy in the presence of axion. The 00 component of the energy-momentum tensor $T^{00}$ in terms of the gauge fields is
\begin{eqnarray}
     T^{00}&=2(D^0\Phi^\dagger)(D^0\Phi)-(W^a)^{0i}(W^a)^0_{~i}-B^{0i}B^0_{~i}-\frac{1}{2}g_{a\gamma\gamma}a\bigg[2\sin^2\theta_W(\tilde{W}^3)^{0i}(W^3)^0_{~i}\nonumber\\
     &+2\cos^2\theta_W \tilde{B}^{0i}B^0_{~i}+\sin\theta_W\cos\theta_W\left(\tilde{B}^{0i}(W^3)^0_{~i}+(\tilde{W}^3)^{0i}B^0_{~i}\right)\bigg]-\mathcal{L}_{eff}\;.
\end{eqnarray}
The total energy of monopole in the KSVZ model can be obtained as
\begin{eqnarray}
E&=&E_{\rm CMM}+E_{\rm axion~int.}\;,\\
E_{\rm CMM}&=&\frac{4\pi}{g'^2}\int_0^\infty{\frac{dr}{2r^2}}\nonumber \\
&&+4\pi\int_0^\infty dr\Big\{\frac{r^2}{2}\Big(\frac{d\rho}{dr}\Big)^2+\frac{1}{g^2}\Big[\Big(\frac{df}{dr}\Big)^2+\frac{r^2}{2}\Big(\frac{dA}{dr}\Big)^2+\frac{1}{2}\frac{(f^2-1)^2}{r^2}+A^2f^2\Big]\nonumber\\
&&+\frac{r^2}{2g'^2}\Big(\frac{dB}{dr}\Big)^2+{\lambda_\Phi\over 4}r^2(\rho^2-v^2)^2+\frac{1}{4}f^2\rho^2+\frac{r^2}{8}\rho^2(A-B)^2\Big\}\;,\\
E_{\rm axion~int.}&=&2\pi g_{a\gamma\gamma}\left(\frac{\sin^2\theta_W}{g'^2}\right)\int_0^\infty{dr a(r)\left[\left(\frac{dA}{dr}\right)+(1-f^2)\left(\frac{dB}{dr}\right)\right]}\;,
\label{eq:EaCMM}
\end{eqnarray}
where the first term of $E_{\rm CMM}$ is apparently infinite. The other terms in $E_{\rm CMM}$ and $E_{\rm axion~int.}$ are all finite. The origin of this infinite energy comes from the singularity of the point-like magnetic monopole of $U(1)_Y$. In Refs.~\cite{Bae:2002bm,Cho:2012bq,Cho:2013vba,Cho:2019vzo}, the author utilized the UV regularization to include the quantum correction and proved the existence of electroweak monopole with a finite energy. According to the Lagrangian with a non-trivial hypercharge U(1) permittivity $\epsilon(\rho)$ in Eq.~(\ref{eq:LUV}), the modified energy $E_{\rm CMM}$ becomes
\begin{eqnarray}
E'_{\rm CMM}
&=&4\pi\int_0^\infty{dr}\Big\{\frac{1}{2}(r\dot{\rho})^2+\frac{1}{g^2}\left(\dot{f}^2+\frac{(r\dot{A})^2}{2}+\frac{(f^2-1)^2}{2r^2}+f^2A^2\right)+\epsilon(\rho)\frac{(r\dot{B})^2}{2{g'}^2}\nonumber\\
&&+\frac{\lambda_\Phi}{4}r^2(\rho^2-v^2)^2+\frac{f^2\rho^2}{4}+\frac{r^2}{8}(A-B)^2\rho^2\Big\}+4\pi\int_0^\infty{dr\frac{\epsilon(\rho)}{2{g'}^2}\cdot\frac{1}{r^2}}\;,
\label{equ:modified_E}
\end{eqnarray}
where $\dot{X}\equiv dX/dr$.
Given the analytic permittivity $\epsilon(\rho)$ in Eq.~(\ref{eq:alphabeta}), the energy becomes finite at the origin. We evaluate the above integrals to obtain the energy $E$ for the electroweak monopole (with $g_{a\gamma\gamma}=0$ and $a(r)=0$) and that in the presence of axion.

It turns out that this regularization offers better mass convergence compared to the method in Ref.~\cite{Cho:2012bq}. In Ref.~\cite{Cho:2012bq}, a scale-independent coefficient is introduced to counteract the divergence caused by the Dirac magnetic monopole. However, this regularization method has a significant drawback: the total energy of monopole is highly dependent on the choice of the boundary condition $f(0)$. As $f(0)$ increases, the total energy diverges, which leads to scale divergence of the monopole energy.
In contrast, the method we utilized in this work introduces a non-trivial permittivity $\epsilon=\epsilon(\rho)$ which is coordinate dependent. This regularization not only cancels the energy divergence but also effectively avoids the problem of scale divergence. Here, the scale is the power exponent $n$ in $\epsilon(\rho)=(\rho/\rho_0)^n$.

\begin{figure}[htp!]
\centering
\vspace{0.2cm}
\subfigtopskip=2pt
\subfigbottomskip=2pt
\subfigcapskip=-0pt
\subfigure[]{
\label{a}
\includegraphics[width=0.46\linewidth]{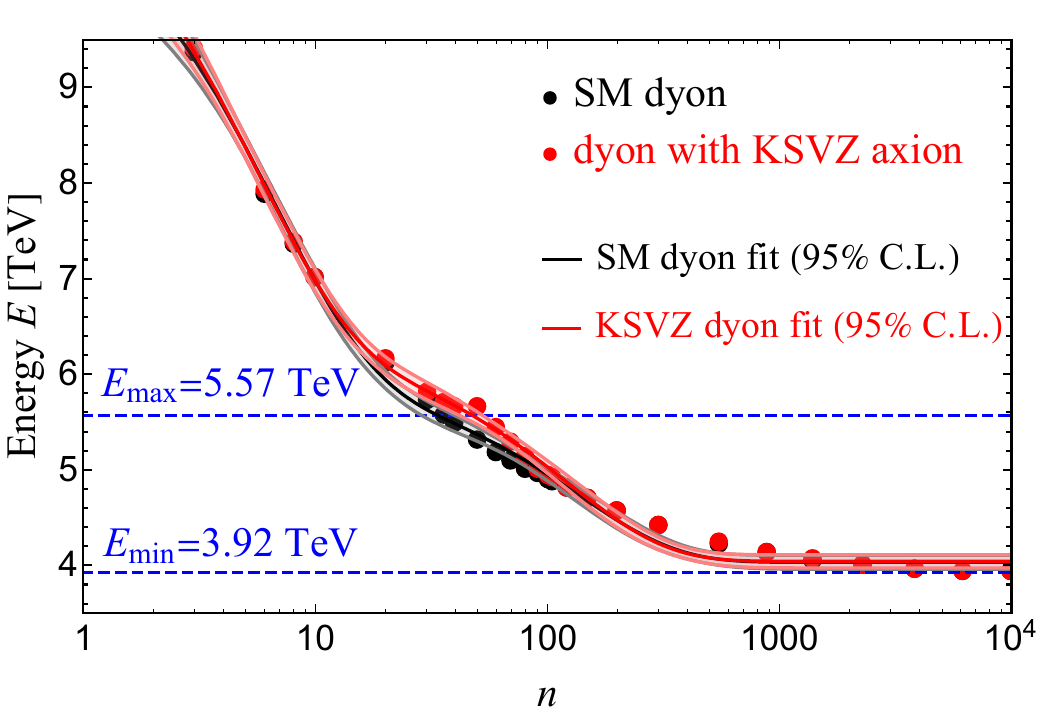}}
\subfigure[]{
\label{b}
\includegraphics[width=0.49\linewidth]{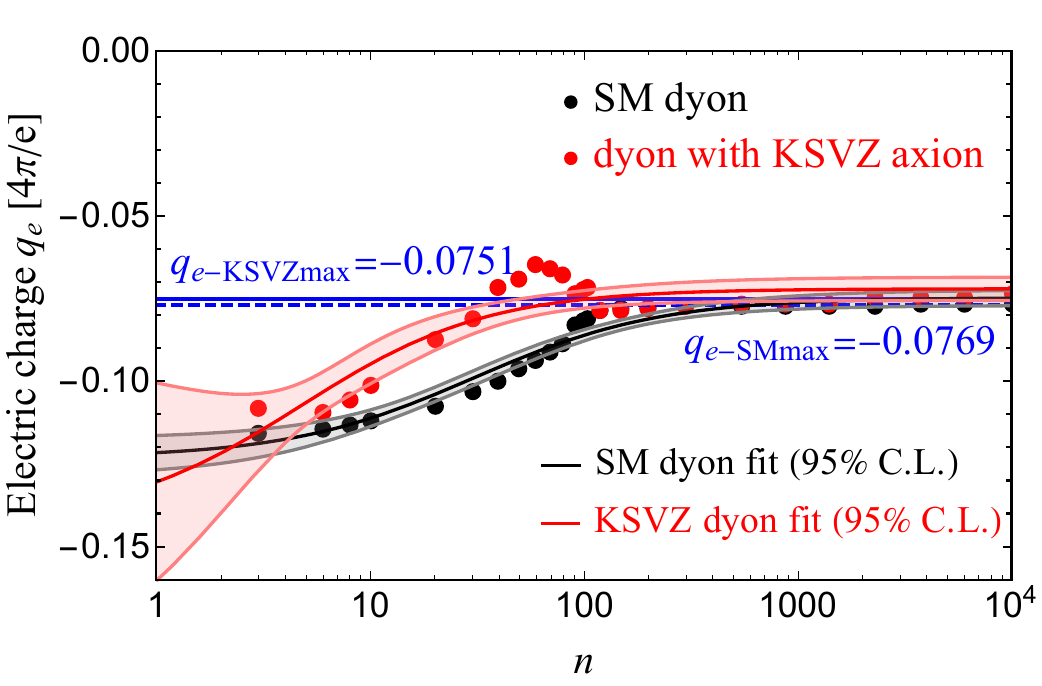}}
\caption{
The total energy (left) and electric charge $q_e$ (right) of the SM dyon (black dot and 95\% C.L. fitted line) and the dyon in the presence of the KSVZ axion (red dot and 95\% C.L. fitted line) as a function of the parameter $n$.
}
\label{fig:numerical}
\end{figure}

The left panel of Fig.~\ref{fig:numerical} shows the total energy (rest mass) of the SM dyon (black dot and 95\% C.L. fitted line) and the dyon in the presence of the KSVZ axion (red dot and 95\% C.L. fitted line) as a function of the parameter $n$. When $n$ takes large values ($n\gg 100$), the contribution of $\epsilon(\rho)$ to the total energy gradually approaches zero, and the lower bound of the dyon mass is then given by the terms in Eq.~\eqref{equ:modified_E} that are independent of $\epsilon$. The numerical calculations show that for $n\geq3$, the mass of SM dyon lies in the range of $3.92~\TeV-9.38~\TeV$, while in the model incorporating the KSVZ axion, the mass range is $3.92~\TeV-9.4~\TeV$.
Moreover, according to Ref.~\cite{Ellis:2016glu}, which is based on experimental value from the Higgs to two-photon decay channel at the LHC, the upper energy bound can be further constrained to $5.57~\TeV$. Within the regularization adopted in this work, this upper energy bound requires the parameter of the SM dyon to be $n\geq 36$.

For some illustrative values of $n$, we show the obtained dyon masses in units of TeV in the third column of Table~\ref{tab:masscharge}.
One can see that, with the presence of axion, the dyon mass is slightly increased within a range of approximately $0.2\%-6\%$.
Moreover, it is noteworthy that the increase in dyon mass arises from two distinct sources: one due to the decrease of $n$, and the other from the introduction of the KSVZ axion. The axion contribution to the mass is from the axion interaction in Eq.~(\ref{eq:EaCMM}) and is governed by $g_{a\gamma\gamma}$ coupling. These two effects are generally coupled. Therefore, to isolate the net energy increase solely contributed by the KSVZ axion, we define the magnitude of the increase as $\Delta E=E^{\rm KSVZ}-E^{\rm SM}$.
In the absence of any experimental constraints, the maximum energy corresponds to $n=3$, at which point the net energy increase contributed by the KSVZ axion to the dyon is $\Delta E=0.026~\TeV$ (increased by 0.27\%). Under the constraint from LHC Higgs to two-photon decay channel, the upper energy bound corresponds to $n=36$, where the net energy increase reaches $\Delta E=0.147~\TeV$ (increased by 2.64\%).

Suppose the electroweak monopole will be observed in future, its property may confirm the nature of axion. On the other hand, the observation of axion would also have impact on the monopole of TeV scale.

\begin{table}[htb!]
\centering
\begin{tabular}{c|c|c|c}
\hline
$n$ value &Characteristics of dyon & mass (in units of TeV) & electric charge (in units of $4\pi/e$) \\
\hline
\multirow{2}{*}{$~n=3~$} &w/o axion & 9.38 & $-0.115$ \\
\cline{2-4}
&w/ axion & 9.40 & $-0.108$\\
\hline
\multirow{2}{*}{$~n=6~$} &w/o axion & 7.88 & $-0.114$ \\
\cline{2-4}
&w/ axion & 7.93 & $-0.109$\\
\hline
\multirow{2}{*}{$~n=36~$} &w/o axion & 5.57 & $-0.101$ \\
\cline{2-4}
&w/ axion & 5.72 & $-0.0737$\\
\hline
\multirow{2}{*}{$~n=50~$} &w/o axion & 5.32 & $-0.0961$ \\
\cline{2-4}
&w/ axion & 5.66 & $-0.0694$\\
\hline
\multirow{2}{*}{$~n=100~$} &w/o axion & 4.91 & $-0.0817$ \\
\cline{2-4}
&w/ axion & 4.96 & $-0.0721$\\
\hline
\end{tabular}
\caption{
The characteristics of dyon without or with the presence of KSVZ axion, including dyon mass in units of TeV (the third column) and electric charge in units of $4\pi/e$ (the fourth column). The results for some illustrative values of $n$ are shown.
}
\label{tab:masscharge}
\end{table}

\subsection{The electromagnetic charges}

We next discuss the electromagnetic properties of electroweak monopole in the presence of KSVZ axion. According to the ansatz of neutral gauge components in Eq.~(\ref{equ:CMM_solution}), we can obtain the electromagnetic tensor $F_{\mu\nu}$ given by the electromagnetic field $A_\mu$ in Eq.~(\ref{eq:Amu}).
The corresponding electric field $\vec E$ and magnetic field $\vec B$ can be then written as
\begin{eqnarray}
\vec{E}&=&-e\left(\frac{\dot A(r)}{g^2}+\frac{\dot B(r)}{g'^2}\right)\hat{r}\;,\\
\vec{B}&=&-\frac{\hat{r}}{er^2}\;.
\end{eqnarray}
where the extra minus sign on the right-handed side of $\vec{B}$ is due to our choice of metric tensor ${\rm diag}(+1,-1,-1,-1)$.
These electromagnetic fields are purely generated by the electroweak monopole in vacuum. However, if we take into account the axion which couples to the physical photon, the axion acts as a kind of source to excite another electromagnetic field.
It changes the electric field as
\begin{eqnarray}
\vec{E}\to \vec{E}+\vec{E}_a = \vec{E}+ g_{a\gamma\gamma}a \vec{B}\;.
\end{eqnarray}
For the electric and magnetic charges of the electroweak dyon, they are defined as the integrals of electromagnetic fields over a closed curve.
We have the following charges
\begin{eqnarray}
q_e&=&\oint_{r=\infty}{d\vec S\cdot(\vec{E}+g_{a\gamma\gamma}a \vec{B})}=-4\pi e\int_0^\infty{dr\left[r^2\left(\frac{\dot A(r)}{g^2}+\frac{\dot B(r)}{g'^2}\right)\right]^\prime}+g_{a\gamma\gamma}\oint_{r=\infty}{d\vec S\cdot a\vec B}\nonumber \\
&=&-4\pi e\left[r^2\left(\frac{\dot A(r)}{g^2}+\frac{\dot B(r)}{g'^2}\right)\right]\Bigg{|}_{r=\infty}+q_m g_{a\gamma\gamma}a(r)\Bigg{|}_{r=\infty}\nonumber\\
&=&-\frac{8\pi}{e}\sin^2{\theta_W}\int_0^\infty{dr f^2(r)A(r)}+\frac{4\pi}{e}\int_0^\infty{dr r^2 T_A}+q_m g_{a\gamma\gamma} a(\infty)\;,\\
q_m&=&\oint_{r=\infty}{d\vec S\cdot\vec B}=-\frac{4\pi}{e}\;.
\label{equ:def1}
\end{eqnarray}
The magnetic charge $q_m$ is exactly the same as that of the Cho-Maison monopole. The first term of electric charge $q_e$ is that of the pure Cho-Maison monopole and the last two terms are induced by the presence of axion-photon interaction.

We show the modified electric charge in the right panel of Fig.~\ref{fig:numerical} and in the fourth column of Table~\ref{tab:masscharge} in units of $4\pi/e$. One can see that the introduction of KSVZ axion exhibits a non-monotonic influence on the electric charge of dyon.
As $n$ increases, the presence of axion leads to an increase in the electric charge with $\Delta q_e=q_e^{\rm KSVZ}-q_e^{\rm SM}>0$.
Given the presence of KSVZ axion, the magnitude of electric charge $|q_e|$ is at most changed by $31.2\%$ for $n=60$ on the 95\% C.L. fitted line. Given $n=36$, the change of the magnitude of electric charge is 27.2\%.
For $n\gg 100$, both electric charges achieve constant values with $q_e^{\rm SM}=-0.0769 {4\pi\over e}$ and $q_e^{\rm KSVZ}=-0.0751 {4\pi\over e}$.
The constant difference purely arises from Witten effect term $q_m g_{a\gamma\gamma} a(\infty)$.

\subsection{The axion potential energy}

\begin{figure}[htb!]
\centering
\includegraphics[scale=1,width=0.45\linewidth]{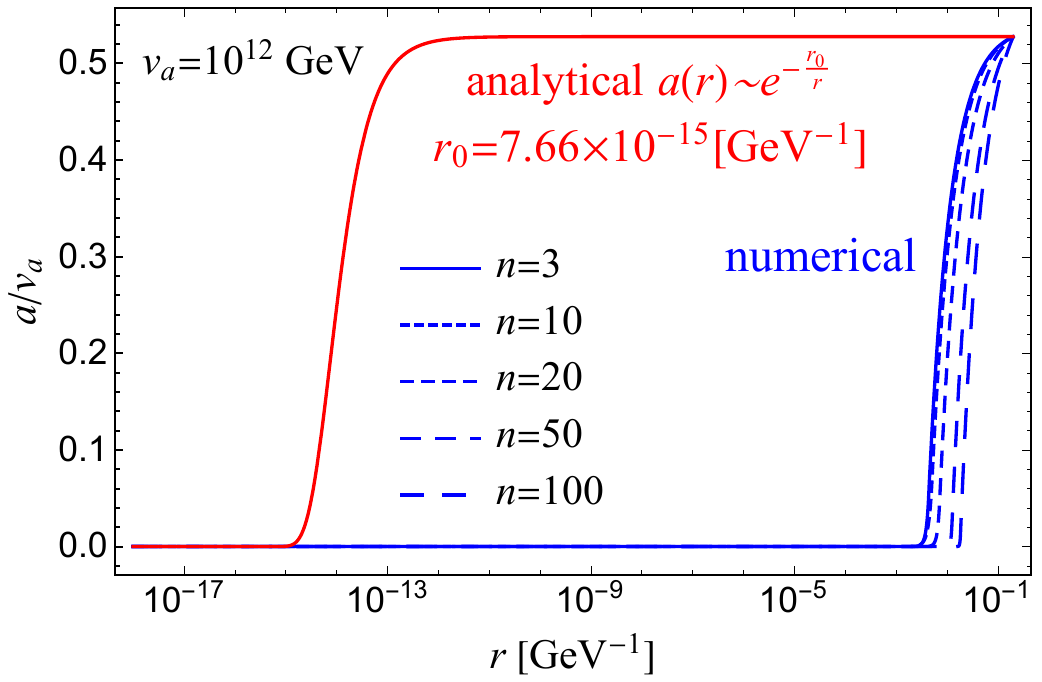}
\includegraphics[scale=1,width=0.49\linewidth]{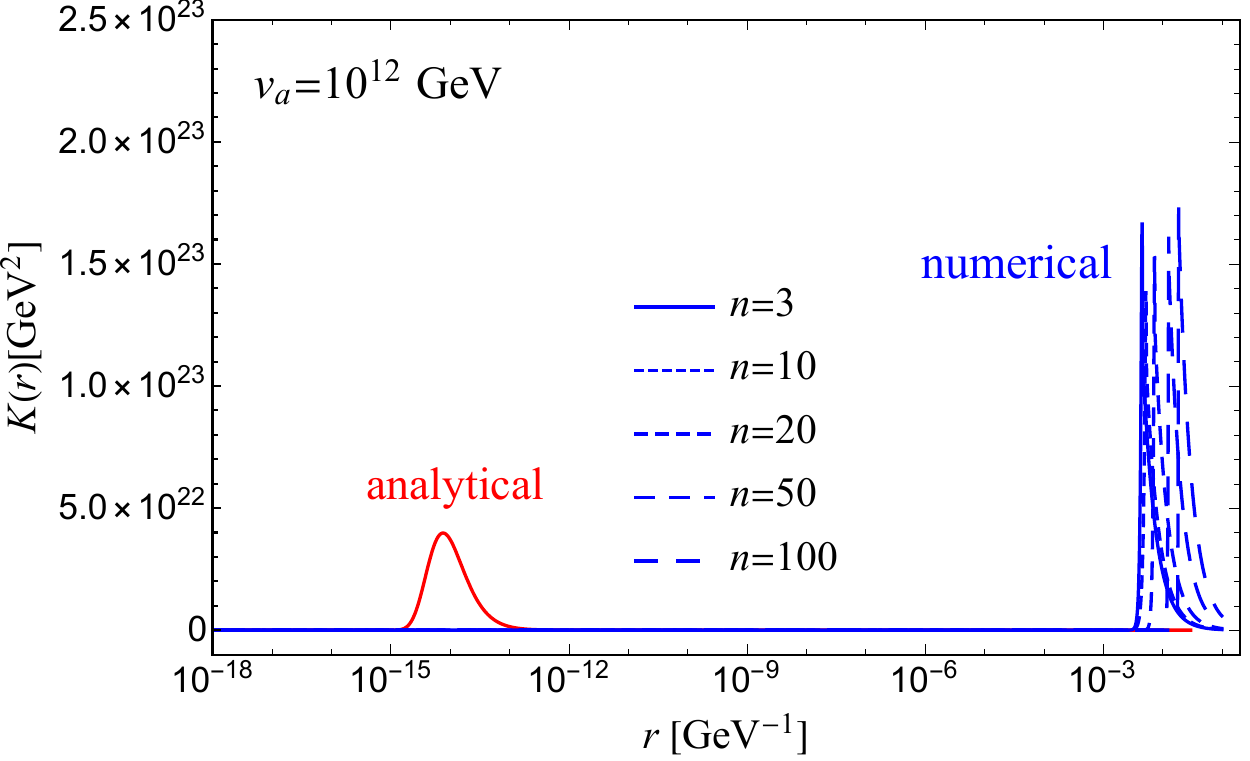}
\includegraphics[scale=1,width=0.45\linewidth]{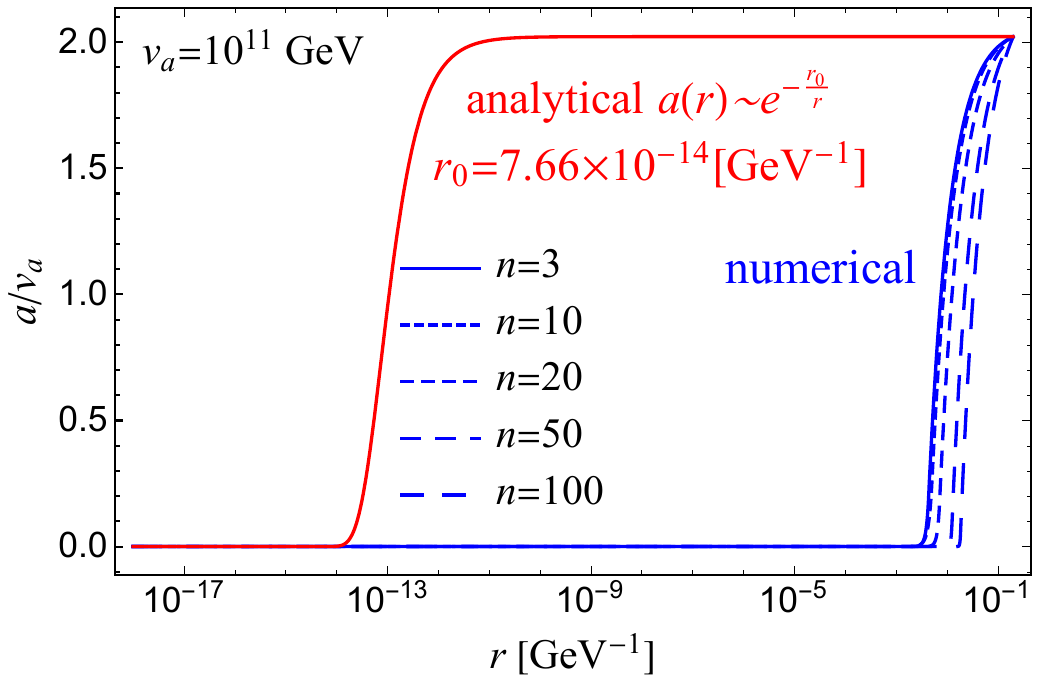}
\includegraphics[scale=1,width=0.49\linewidth]{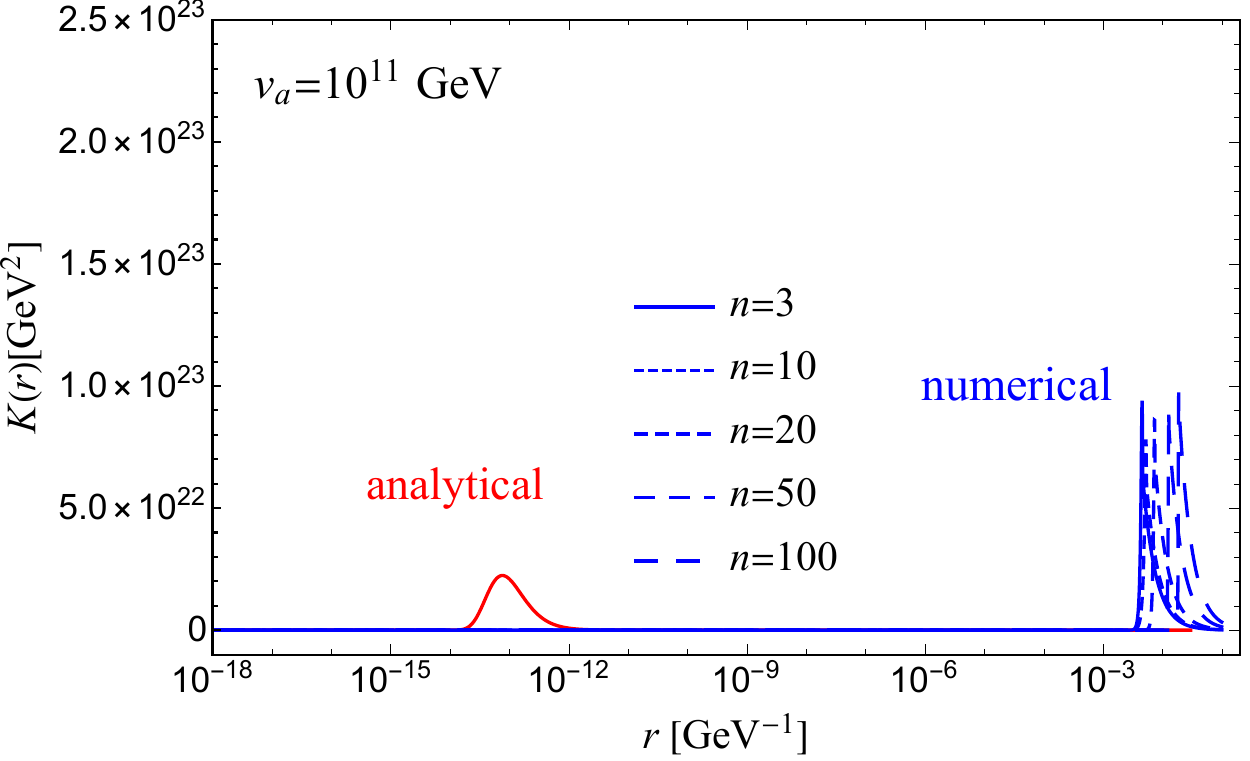}
\caption{The normalized axion field $a/v_a$ (left) and the integrand $K(r)$ in Eq.~(\ref{equ:axion_potential}) (right), as a function of $r$ in units of $\GeV^{-1}$. We compare the result of the analytical form $a(r)\sim e^{-r_0/r}$ (red line) with that of our numerical solution (blue line). Both cases of $v_a=10^{12}$ GeV (top) and $10^{11}$ GeV (bottom) are shown.}
\label{fig:comparison}
\end{figure}

Finally, we reexamine the axion potential in the background field of a monopole which was first studied by Fischler et al. in Ref.~\cite{Fischler:1983sc}. The axion potential energy can be divided into two parts. One part is provided by the kinetic energy of axion. The axion itself has a lot of kinetic energy. The other part comes from the electrostatic field energy $1/2\vec E_a^2$. The induced electric field $\vec E_a$ is proportional to the magnetic field of monopole $\vec{B}$. Thus, there exists a non-negligible energy cost $1/2\vec E_a^2$ in the vicinity of the axion.
In other words, the axion is strongly repelled by the monopole in electromagnetic properties and needs to pay a price of large energy if it wants to be combined with monopole. The axion potential energy is given by
\begin{eqnarray}
V_a&=&\frac{1}{2}\int{dV\left[\left(\nabla a\right)^2+\vec E_a^2\right]}\nonumber\\
&=&E_{\rm axion~kin.}+\frac{1}{2}\int{dV\left[g_{a\gamma\gamma}^2\frac{a(r)^2}{e^2r^4}\right]}\nonumber\\
&=&{a(\infty)\over v_a}\int_0^\infty{drr^2\left(\frac{da}{dr}\right)^2}+{a(\infty)\over v_a}\int_0^\infty{dr \left(\frac{g_{a\gamma\gamma}a(r)}{e r}\right)^2}\equiv \int_0^\infty dr K(r)\;.
\label{equ:axion_potential}
\end{eqnarray}
Following Eq.~\eqref{eq:axion_boundary}, it is convenient to introduce the dimensionless asymptotic boundary parameter $\theta_\infty \equiv a(\infty)/v_a$.
We can then follow the substitution of integral variable in Ref.~\cite{Fischler:1983sc} to minimize $V_a$. The axion potential can be rewritten as
\begin{eqnarray}
V_a=-\theta_\infty\left[r_0\int_0^\infty{dz\left(\frac{da}{dz}\right)^2}+\frac{g_{a\gamma\gamma}^2}{e^2}\frac{1}{r_0}\int_0^\infty{dza^2}\right]\;,
\end{eqnarray}
where the radius $r$ is replaced with a dimensionless variable $z=r_0/r$ with $r_0=|g_{a\gamma\gamma}/e|$.
The axion in the configuration of $a(r)=a(\infty) e^{-r_0/r}$ approaches $a(\infty)$ when $z\to 0$ and can minimize $V_a$~\cite{Fischler:1983sc}. The minimum potential energy is then given by
\begin{eqnarray}
V_a^{\rm min}=-\frac{a(\infty)^3}{2v_a}\left(r_0+\frac{g_{a\gamma\gamma}^2}{e^2r_0}\right)=-{a(\infty)^3\over v_a}\frac{g_{a\gamma\gamma}}{e}=-\theta_\infty^2\frac{c_{a\gamma\gamma}}{e}a(\infty)\;.
\label{equ:analytical_Va}
\end{eqnarray}
The above result is equivalent to that of Ref.~\cite{Fischler:1983sc}.
Note that the estimate of $V_a^{\rm min}$ should be regarded only as the minimum energy associated with the classical asymptotic axion configuration, rather than the numerical calculation of the embedded KSVZ axion solution.
As seen from the left panels of Fig.~\ref{fig:comparison}, the analytical profile $a(r)\sim e^{-r_0/r}$ (red line) agrees with our numerical solution of axion field (blue line) only for $r\gg r_0$.
For small radial distance, there exhibits significant distinction between them. In the classical profile, the axion field starts to rise already at very short distance, whereas in the numerical solution the axion field remains suppressed near the monopole core and rises only at somewhat larger $r$. This delayed growth pattern is the consequence of dynamically embedding the KSVZ axion into the electroweak monopole background. As $v_a$ decreases, the analytical result approaches the numerical solution. The integrand $K(r)$ is shown in the right panels of Fig.~\ref{fig:comparison}. Since the energy density is highly sensitive to the short distance behavior of the axion profile, this different property in the core region leads to a sizable distinction in the total energy.
In Table~\ref{tab:axion_potential}, we show the comparison of the axion potential energy $V_a$ between the minimum value calculated as above and the one obtained by integrating the numerical solution of the axion field. In particular, the numerical solution yields a much larger integrated value of $V_a$. This is because the dominant contribution from the integrand $K(r)$ is shifted to larger radii than that in the analytical configuration. Physically, this implies that the monopole background suppresses the axion field near the central core. As a result, the dominant contribution to the axion energy is pushed outward to larger radii.

\begin{table}[htb!]
\centering
\begin{tabular}{|c|c|c|c|c|c|c|c|}
\hline
$v_a[\GeV]$ &$V_a^\text{min}[\GeV]$ & \multicolumn{6}{c|}{\text{numerical result }$V_a[\GeV]$} \\ \hline
\multirow{2}{*}{$10^{12}$} & \multirow{2}{*}{$1.13\times 10^{9}$} & $n$ & 3 & 10 & 20 & 50 &100\\ \cline{3-8}
                   &                    & $V_a$ & $5.72\times 10^{20}$ & $2.93\times 10^{20}$ & $6.55\times 10^{20}$ & $1.85\times 10^{21}$ & $2.90\times 10^{21}$\\ \hline
\multirow{2}{*}{$10^{11}$} & \multirow{2}{*}{$6.33\times 10^9$} & $n$ & 3 & 10 & 20 & 50 &100\\ \cline{3-8}
                   &                    & $V_a$ & $3.22\times 10^{20}$ & $1.65\times 10^{20}$ & $3.68\times 10^{20}$ & $1.04\times 10^{21}$ & $1.63\times 10^{21}$ \\ \hline
\end{tabular}
\caption{The comparison of the axion potential in Eq.~\eqref{equ:axion_potential} between the minimum value calculated from the analytical formula of axion solution in Eq.~(\ref{equ:analytical_Va}) and the one by integrating the numerical solution of the axion field from EoM.
}
\label{tab:axion_potential}
\end{table}

\section{Conclusion}
\label{sec:Con}

The Witten effect implies the dynamics of axion and magnetic monopole. The Cho-Maison monopole is a realistic electroweak monopole arisen in the Weinberg-Salam theory. This monopole of TeV scale mass motivates the dedicated search for electroweak monopole at colliders. It is plausible to explore the implication of axion for the electroweak magnetic monopole.

In this work we investigate the topological solutions of Cho-Maison electroweak monopole in the presence of KSVZ axion. We use the spherically symmetric ansatz for the electroweak monopole and introduce the spherically symmetric function for the axion field. The effective Lagrangian is then showed in terms of the radial functions. It includes the electroweak monopole part, the axion kinetic energy as well as the axion interaction term. We derive the consequent equations of motion in the presence of the axion-photon coupling and show the numerical results. Given the above topological solutions, we calculate the characteristics of the electroweak monopole such as the monopole mass and the electromagnetic charges. A non-trivial hypercharge $U(1)$ permittivity-like function is used for UV regularization.

We find that the monopole mass can be slightly changed by the axion-photon interaction. The presence of KSVZ axion changes the electric charge of the monopole by 30\% at the given PQ scale $v_a=10^{12}~\text{GeV}$. These changes would have impacts on either the existence of axion or the testability of electroweak monopole. We also calculate the axion potential energy in terms of the axion solution from EoM, and compare with the result in the classical limit.

\acknowledgments
We would like to thank Yongcheng Wu and Ke-Pan Xie for very useful discussions. T.~L. is supported by the National Natural Science Foundation of China (Grant No. 12375096, 12035008, 11975129).

\bibliography{refs}

\end{document}